\documentclass[letterpaper,twocolumn,american,nofootinbib]{revtex4}
\usepackage{ae}
\usepackage{aecompl}
\usepackage[T1]{fontenc}
\usepackage[latin1]{inputenc}
\usepackage{amsmath}
\usepackage{graphicx}
\usepackage{amssymb}

\makeatletter

\providecommand{\tabularnewline}{\\}

\usepackage{psfrag}

\usepackage{babel}
\makeatother
\begin{document}

\title{Cosmological particle production and the precision of the WKB approximation }

\author{Sergei Winitzki}

\affiliation{Department of Physics, Ludwig-Maximilians University, 80333 Munich,
Germany}

\date{\today}

\begin{abstract}
Particle production by slow-changing gravitational fields is usually
described using quantum field theory in curved spacetime. Calculations
require a definition of the vacuum state, which can be given using
the adiabatic (WKB) approximation. I investigate the best attainable
precision of the resulting approximate definition of the particle
number. The standard WKB ansatz yields a divergent asymptotic series
in the adiabatic parameter. I derive a novel formula for the optimal
number of terms in that series and demonstrate that the error of the
optimally truncated WKB series is exponentially small. This precision
is still insufficient to describe particle production from vacuum,
which is typically also exponentially small. An adequately precise
approximation can be found by improving the WKB ansatz through perturbation
theory. I show quantitatively that the fundamentally unavoidable imprecision
in the definition of particle number in a time-dependent background
is equal to the particle production expected to occur during that
epoch. The results are illustrated by analytic and numerical examples. 
\end{abstract}
\maketitle

\section{Introduction}

Particle production by gravity in a slowly expanding universe can
be described using quantum field theory in curved spacetime (QFTCS)~\cite{DeWitt75,BirDav82,Ful89,GriMamMos94,For97,Jac03,MukWin05}.
A well-known feature of QFTCS is the absence of an absolute definition
of vacuum and particles for quantum fields in arbitrary curved backgrounds
(see e.g.~\cite{Dav84}). The number of particles detected by an
observer depends on the observer's motion, and there are no preferred
observers in a general spacetime. If the universe is sufficiently
spatially flat and expands sufficiently slowly so that the four-curvature
scale is much larger than the wavelength of a field mode, there is
a natural (if approximate) definition of the vacuum state for that
mode: the adiabatic vacuum with respect to a given fiducial time $t_{0}$.
This is the vacuum state seen by {}``approximately inertial'' observers
at $t=t_{0}$. However, an adiabatic vacuum state defined at $t=t_{0}$
is generally an excited state with respect to the vacuum defined at
another time $t_{1}\neq t_{0}$. Such field excitations are interpreted
as particles produced by gravity. Since there is no single physically
preferred vacuum state, we speak of an \textsl{apparent} particle
production. The {}``real'' particle content of a given quantum state
of the field cannot be unambiguously established without referring
to a particular physical experiment where the vacuum state is observed
or prepared.

For instance, the most straightforward definition of the vacuum state
(the {}``instantaneous diagonalization'' of the Hamiltonian) yields
an infinite apparent particle production in some generic cases~\cite{Ful79},
while the adiabatic vacuum exhibits finite particle densities in the
same cases. It has been also proposed~\cite{Par69} that the vacuum
state should be chosen to minimize the apparent particle number observed
at a time $t_{0}$. This prescription is physically reasonable and
yields a quantum state close to an adiabatic vacuum, but the resulting
state will in general depend on the choice of $t_{0}$. A natural
and unambiguous definition of the vacuum state is available only in
regimes where gravity becomes negligible. So the concept of particles
can be used in curved spacetimes only in an approximate sense which
becomes more precise in slow-changing, almost flat geometries, and
also for high-energy particles. The central theme of this paper is
to explicitly analyze the precision of this approximation.

The prescription of the adiabatic vacuum is based on the WKB approximation
and has been particularly useful in the context of QFTCS (see e.g.~\cite{Par69,ZelSta71,BunChrFul78}).
It is well known that the WKB approximation is applicable to equations
such as\begin{equation}
\frac{d^{2}x}{dt^{2}}+\omega^{2}(t)x=0,\label{eq:osc equ T}\end{equation}
where $\omega(t)>0$ is a time-dependent frequency function. The approximate
solutions can be found in the form of an asymptotic series in the
adiabatic parameter $T^{-1}$, where $T$ is the characteristic variation
timescale of the function $\omega(t)$, which is assumed to be slow-changing~\cite{BenOrs78}.
Using the $n$-th order WKB approximation, one can define adiabatic
vacuum states of order $n$~\cite{BunChrFul78}. The difference between
$n$-th order adiabatic vacua defined at different fiducial times
is characterized by apparent particle occupation numbers, which are
of order $T^{-1-n}$; this is then the imprecision in the resulting
definition of particle numbers. A brief overview of the calculation
of cosmological particle production using the adiabatic vacuum prescription
is given in Sec.~\ref{sub:Quantum-fields-in}. 

It is known that the WKB approximation involves a divergent asymptotic
series%
\footnote{I shall show in Sec.~\ref{sub:Divergence-of-the} that the WKB series
generally diverges. Although this statement appears to be common knowledge,
I was unable to find a derivation in the literature. A closely related
result is the Borel summability of asymptotic series for adiabatic
invariants~\cite{CosDupKru04}; see also Ref.~\cite{DunHal99}.%
} and thus cannot be used beyond a certain order $n_{\max}$. Therefore,
particle occupation numbers computed with respect to an adiabatic
vacuum are defined only with a certain fundamentally limited accuracy.
If one uses the adiabatic vacuum of the optimal order $n_{\max}$,
one obtains particle occupation numbers up to an uncertainty of order
$T^{-1-n_{\max}}$, and this accuracy cannot be improved any further. 

Since typical particle numbers produced in vacuum by smooth geometries
are exponentially small~\cite{BirDav82}, namely of order $\exp(-\omega T)$,
it is \emph{a priori} unclear whether even the best attainable precision
$\sim T^{-1-n_{\max}}$ is adequate for the description of particle
production. The exponentially small particle numbers can be calculated
if one applies perturbation theory techniques to the WKB ansatz; various
such techniques are outlined in Sec.~\ref{sec:Perturbative-improvement-of}.
I shall summarily refer to these techniques as the \textsl{perturbatively
improved WKB}. 

Although the WKB expansion, being a power series in $T^{-1}$, necessarily
misses any exponentially small terms, the quantity $T^{-n_{\max}}$
\emph{can} be of order $\exp(-\omega T)$ if $n_{\max}$ is sufficiently
large, say of order $T$. The main result of the present work is an
explicit estimate of the optimal order $n_{\max}$ and of the resulting
optimal precision of the WKB series. The current literature does not
appear to offer such direct estimates. It is difficult to analyze
the WKB series since there is no closed-form expression for the $n$-th
term of that series. To circumvent this difficulty, I use a particular
perturbatively improved WKB technique---the so-called Bremmer series
(see Sec.~\ref{sub:The-Bremmer-series}). I show in Sec.~\ref{sec:Properties-of-the}
that $n_{\max}=O(\omega T)$ and the error of the optimally truncated
series is exponentially small. More precisely, for Eq.~(\ref{eq:osc equ T})
with analytic functions $\omega(t)$, the optimal order $n_{\max}$
of the WKB approximation at $t=t_{0}$ is found as\begin{equation}
n_{\max}=\min_{t_{i}}\left|\int_{t_{0}}^{t_{i}}\omega(t)dt\right|,\label{eq:nmax def}\end{equation}
where the complex numbers $t_{i}$, $i=1,2,...$ are all the zeros
and the poles of $\omega(t)$ in the complex $t$ plane, and the integrals
are performed along the paths from $t_{0}$ to $t_{i}$ that give
the smallest value to the above integrals. The error of the optimally
truncated WKB series is of order\begin{equation}
n_{\max}^{-1/2}\exp\left(-2n_{\max}\right).\label{eq:error nmax}\end{equation}
 I also show that the smallest attainable uncertainty in the definition
of particle numbers during a time-dependent epoch is of the order
of the particle production expected to occur within that epoch. These
issues are discussed in Sec.~\ref{sub:Particle-production-and}.
I illustrate these estimates by analytic and numerical examples in
Sec.~\ref{sec:Examples}.

\section{Adiabatic approximation in QFTCS}

\subsection{The WKB approximation\label{sub:The-WKB-approximation}}

The WKB approximation, also known in the mathematical literature as
the phase integral method and the Liouville-Green approximation~\cite{Hea62,Olv74},
applies to equations of the form~(\ref{eq:osc equ T}), rewritten
as\begin{equation}
\varepsilon^{2}\frac{d^{2}x}{dt^{2}}+\omega^{2}(t)x=0,\label{eq:osc equ}\end{equation}
where $\varepsilon$ is a formal parameter (we shall set $\varepsilon=1$
at the end of all calculations). The frequency $\omega(t)$ is assumed
to be a sufficiently slow-varying function of time, so that the adiabaticity
condition\begin{equation}
\varepsilon\left|\frac{d\omega}{dt}\right|\ll\omega^{2}\label{eq:adiab cond}\end{equation}
 holds for all values of time $t$ to be considered. We shall additionally
assume throughout this paper that $\omega^{2}(t)>0$ for all relevant
$t$, and that $\omega(t)$ is an analytic function. The well-known
WKB ansatz is \begin{equation}
x_{\textrm{WKB}}(t)=\frac{C}{\sqrt{\omega(t)}}\exp\left(\pm\frac{i}{\varepsilon}\int^{t}\omega(t')dt'\right).\label{eq:WKB 0}\end{equation}
 The error of this approximation is of order $\varepsilon^{2}$ if
$\omega(t)$ is a sufficiently well-behaved function~\cite{Hea62,Olv74}.
One can generalize the ansatz~(\ref{eq:WKB 0}) to an $n$-th order
approximation (see e.g.~\cite{Kul57,Cha73,BenOrs78}),\begin{equation}
x_{\textrm{WKB}}^{(n)}(t)=\frac{C}{\sqrt{W_{n}(t)}}\exp\left(\pm\frac{i}{\varepsilon}\int^{t}W_{n}(t')dt'\right),\end{equation}
 where the auxiliary function $W_{n}(t;\varepsilon)$ is an approximate
solution of \begin{equation}
\frac{1}{2}\frac{W^{\prime\prime}}{W}-\frac{3}{4}\frac{W^{\prime2}}{W^{2}}=\frac{\omega^{2}-W^{2}}{\varepsilon^{2}}\label{eq:W equ}\end{equation}
which is found as a power series in $\varepsilon^{2}$, \begin{align}
W_{n} & \equiv\sum_{k=0}^{n}\varepsilon^{2k}S_{k}(t)=\omega-\varepsilon^{2}\left(\frac{1}{4}\frac{\ddot{\omega}}{\omega^{2}}-\frac{3}{8}\frac{\dot{\omega}^{2}}{\omega^{3}}\right)\label{eq:W series}\\
\negmedspace+\varepsilon^{4} & \negmedspace\left(\frac{1}{16}\frac{\omega^{(4)}}{\omega^{4}}-\frac{5}{8}\frac{\dddot{\omega}\dot{\omega}}{\omega^{5}}-\frac{13}{32}\frac{\ddot{\omega}^{2}}{\omega^{5}}+\frac{99}{32}\frac{\ddot{\omega}\dot{\omega}^{2}}{\omega^{6}}-\frac{297}{128}\frac{\dot{\omega}^{3}}{\omega^{7}}\right)\!+...\nonumber \end{align}
The approximation $x_{\textrm{WKB}}^{(n)}(t)$ is accurate up to error
terms of order $O(\varepsilon^{2n+1})$, so the series~(\ref{eq:W series}),
whether convergent or not, is an asymptotic expansion at $\varepsilon\rightarrow0$~\cite{BenOrs78}.
The consecutive terms of the series~(\ref{eq:W series}) can be found
iteratively, for instance, by expanding the RHS of the relation\begin{equation}
W_{n+1}+O\left(\varepsilon^{2n+4}\right)=\sqrt{\omega^{2}-\varepsilon^{2}\left[\frac{1}{2}\frac{W_{n}^{\prime\prime}}{W_{n}}-\frac{3}{4}\frac{W_{n}^{\prime2}}{W_{n}^{2}}\right]}\label{eq:W recur}\end{equation}
 in powers of $\varepsilon^{2}$ up to order $O(\varepsilon^{2n+2})$.
It is clear that $W_{n}(t;\varepsilon)$ is a rational function of
$\omega(t)$ and its derivatives up to the order $2n$. No closed-form
expression appears to be available for the $n$-th term $S_{n}(t)$
of the series~(\ref{eq:W series}).%
\footnote{See, however, Ref.~\cite{BenOlaWan77} for some attempts to simplify
the WKB terms by eliminating total derivatives, Ref.~\cite{RobRom00}
for an approach to make high-order WKB calculations numerically more
manageable, and Ref.~\cite{KudVan02} for an approximate resummation
using Airy functions.%
}

The WKB expansion can be equivalently parametrized by the characteristic
timescale $T$ of the variation of $\omega(t)$. After rescaling the
time variable by $t=T\tau$, the WKB series becomes a power series
in $T^{-1}$. However, we shall not use this parametrization.

\subsection{Quantum fields in expanding universe\label{sub:Quantum-fields-in}}

In this section I briefly review the computational procedure for determining
the (apparent) particle occupation numbers in QFTCS, following~\cite{MukWin05}.
To be specific, I consider a minimally coupled massive scalar field
$\phi(x)$ in a Friedmann-Robertson-Walker universe with flat spatial
sections and the line element $d\tau^{2}-a^{2}(\tau)d\mathbf{x}^{2}$,
where $a(\tau)$ is the scale factor assumed to be a known function
of time $\tau$. It is convenient to pass to the conformal time $t\equiv\int^{\tau}a^{-1}(\tau)d\tau$
(which is denoted by $t$ for consistency with the previous notation)
and to rescale the field $\phi$ as $\phi(x)=a^{-1}\chi(x)$. The
auxiliary field $\chi(x)$ is quantized using a mode expansion of
the form\begin{equation}
\hat{\chi}(t,\mathbf{x})=\int\frac{d^{3}\mathbf{k}}{\left(2\pi\right)^{3/2}}\frac{1}{\sqrt{2}}\left(\hat{a}_{\mathbf{k}}e^{i\mathbf{k}\cdot\mathbf{x}}v_{k}(t)+H.c.\right),\end{equation}
where $\hat{a}_{\mathbf{k}}$ are the annihilation operators, $v_{k}(t)$
are mode functions for the wavenumber $\mathbf{k}$, and {}``$H.c.$''
denotes the Hermitian conjugate terms. The mode functions $v_{k}(t)$
are complex-valued solutions of the equation\begin{equation}
v_{k}^{\prime\prime}+\left(k^{2}+m^{2}-\frac{a^{\prime\prime}}{a}\right)v_{k}\equiv v_{k}^{\prime\prime}+\frac{\omega_{k}^{2}(t)}{\varepsilon^{2}}v_{k}=0,\label{eq:vk equ}\end{equation}
subject to the normalization condition\begin{equation}
\textrm{Im}\left(v_{k}v_{k}^{*\prime}\right)=1,\label{eq:vk norm}\end{equation}
where $m$ is the mass of the field $\phi$ and the prime $^{\prime}$
denotes derivatives with respect to $t$. (In a flat spacetime, we
would have $\omega_{k}(t)=\textrm{const}$ and $v_{k}\sim e^{-i\omega_{k}t/\varepsilon}$.)
Due to Eq.~(\ref{eq:vk norm}), the creation and annihilation operators
satisfy the standard commutation relations, $[\hat{a}_{\mathbf{k}},\hat{a}_{\mathbf{k}'}^{\dagger}]=\delta(\mathbf{k}-\mathbf{k}')$.
The vacuum state $\left|0\right\rangle $ of the field is defined
as usual by\begin{equation}
\hat{a}_{\mathbf{k}}\left|0\right\rangle =0\;\textrm{ for all }\mathbf{k}.\end{equation}
Thus defined, the vacuum state depends on the choice of the mode functions
$v_{k}(t)$. Because of the freedom to multiply each mode by a constant
phase factor, solutions of Eqs.~(\ref{eq:vk equ})-(\ref{eq:vk norm})
may be effectively parametrized by the (complex-valued) ratio $v_{k}^{\prime}/v_{k}$
at a fixed time $t=t_{0}$. Different choices of this ratio yield
mode functions $v_{k}(t)$ describing different vacuum states. 

We shall now focus on the behavior of one field mode with a fixed
$\mathbf{k}$ and hence drop the subscript $k$. The next step is
to apply the WKB approximation to Eq.~(\ref{eq:vk equ}). The $n$-th
order WKB approximation to the mode function is\begin{equation}
v_{\textrm{WKB}}(t)=\frac{\sqrt{\varepsilon}}{\sqrt{W_{n}(t)}}\exp\left(-\frac{i}{\varepsilon}\int^{t}W_{n}(t')dt'\right),\end{equation}
where the factor $\sqrt{\varepsilon}$ ensures that the normalization
condition~(\ref{eq:vk norm}) holds. Using the function $v_{\textrm{WKB}}(t)$,
one defines the adiabatic vacuum of order $n$ at a fiducial time
$t=t_{0}$ by requiring that the mode function $v(t)$ should match
the WKB expression at $t=t_{0}$, namely\begin{equation}
\frac{v'(t_{0})}{v(t_{0})}=\frac{v_{\textrm{WKB}}^{\prime}(t_{0})}{v_{\textrm{WKB}}(t_{0})}.\label{eq:adiab init cond}\end{equation}
 Let us denote the resulting mode function by $v_{0}(t)$. (More generally,
one may require that the matching in Eq.~(\ref{eq:adiab init cond})
should hold only up to terms of order $\varepsilon^{2n+2}$, i.e.
up to the precision of the $n$-th order WKB approximant, but we shall
not make use of this additional freedom.)

The adiabatic vacuum prescription can be applied at a different fiducial
time $t=t_{1}$, yielding another mode function $v_{1}(t)$. These
two mode functions are related by a Bogolyubov transformation\begin{equation}
v_{0}(t)=\alpha v_{1}(t)+\beta v_{1}^{*}(t).\end{equation}
 It is well known that the $t=t_{0}$ vacuum appears to have the number
density $\left|\beta\right|^{2}$ of particles with respect to the
$t=t_{1}$ vacuum. Once the functions $v_{0}(t)$ and $v_{1}(t)$
are found, the Bogolyubov coefficient $\beta$ can be computed as\begin{equation}
\beta=\frac{v_{0}^{\prime}(t)v_{1}(t)-v_{0}(t)v_{1}^{\prime}(t)}{2i}.\label{eq:beta through v}\end{equation}
The r.h.s.~of Eq.~(\ref{eq:beta through v}) is a time-independent
Wronskian and thus can be evaluated at arbitrary $t$, say at $t=t_{1}$.
Since the values of $v_{1}(t_{1})$ and $v_{1}^{\prime}(t_{1})$ are
known from Eq.~(\ref{eq:adiab init cond}) after replacing $t_{0}$
by $t_{1}$, it remains to compute $v_{0}(t_{1})$ and $v_{0}^{\prime}(t_{1})$.
The latter computation requires solving Eq.~(\ref{eq:vk equ}) through
the interval $[t_{0},t_{1}]$ with initial conditions~(\ref{eq:adiab init cond}).
Note that the $n$-th order WKB approximation to $v_{0}(t)$ identically
satisfies the condition~(\ref{eq:adiab init cond}) for all $t$
and thus cannot be used to determine the particle number at $t=t_{1}$.
The mode function $v_{0}(t)$ must be computed either by a more accurate
analytic method or numerically. In principle, a higher-order WKB approximant
could be used if its precision were adequate, but this is not always
the case, as we shall show in the next subsection. (The value of $\beta$
obtained from a higher-order WKB approximant would be of order $\varepsilon^{2n+2}$
while the correct answer is usually exponentially small.)

\subsection{Super-adiabatic regimes}

\begin{figure}
\begin{center}\psfrag{omega}{$\omega(t)$} \psfrag{o1}{$\omega_1$} \psfrag{o2}{$\omega_0$} \psfrag{t1}{$t_0$} \psfrag{t2}{$t_1$} \psfrag{t}{$t$}\includegraphics[%
  width=3in]{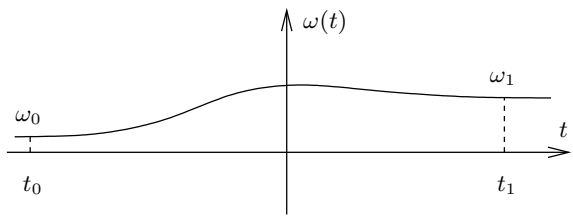}\end{center}

\caption{A frequency function $\omega(t)$ with two super-adiabatic regimes
at $t\leq t_{0}$ and $t\geq t_{1}$.\label{cap:superadiabatic}}
\end{figure}

Let us consider the case where $\omega(t)\approx\textrm{const}$ except
for a finite time interval, for instance $\omega\equiv\omega_{0}$
for $t\leq t_{0}$ and $\omega\equiv\omega_{1}$ for $t\geq t_{1}$
(the values $t_{0}$ and $t_{1}$ may be finite or infinite). A possible
such function $\omega(t)$ is plotted in Fig.~\ref{cap:superadiabatic}.
In that case, the vacuum states are naturally and uniquely defined
for $t\leq t_{0}$ (the {}``in'' vacuum) and for $t\geq t_{1}$
(the {}``out'' vacuum). We shall refer to this situation as an {}``in-out''
transition, and to the regimes $t\leq t_{0}$ and $t\geq t_{1}$ as
super-adiabatic regimes.

More precisely, a super-adiabatic regime at $t=t_{0}$ means that
the adiabaticity condition~(\ref{eq:adiab cond}) becomes an equality,\begin{equation}
\lim_{t\rightarrow t_{0}}\frac{\dot{\omega}}{\omega^{2}}=0,\end{equation}
 and that analogous conditions hold for higher derivatives of $\omega(t)$.
In other words, all derivatives of $\omega(t)$ vanish in a super-adiabatic
regime. The WKB series is truncated, so that we have\begin{equation}
W_{n}(t)=\left\{ \begin{array}{c}
\omega_{0}\textrm{ for }t\leq t_{0},\\
\omega_{1}\textrm{ for }t\geq t_{1},\end{array}\right.\end{equation}
because $W_{n}(t)$ is a local function of $\omega(t)$ and its derivatives
at a point $t$, as seen from Eq.~(\ref{eq:W series}). Thus, within
one super-adiabatic regime, the adiabatic vacuum states of all orders
coincide and yield a natural definition of \emph{the} vacuum state
and an unambiguous notion of particles. For instance, the result of
an {}``in-out'' transition is a well-defined set of particle numbers
(the particle density of the {}``in''-vacuum with respect to the
{}``out''-vacuum). Physically this means that gravity becomes unimportant
in a super-adiabatic regime, and all inertial detectors exactly agree
on the particle content of any quantum state of the field. Outside
of a super-adiabatic regime, adiabatic vacuum states are still well-defined,
but there can be only an approximate agreement between adiabatic vacua
of different orders or defined at different fiducial times.

If the quantum field is in the {}``in''-vacuum state, the particle
numbers observed at $t>t_{1}$ can be unambiguously predicted using
Eq.~(\ref{eq:beta through v}). However, if we used the WKB approximation
to compute the function $v_{0}(t)$ at $t=t_{1}$, we would find $v_{0}(t)\propto e^{i\omega_{1}t/\varepsilon}$,
the Wronskian involved in Eq.~(\ref{eq:beta through v}) would vanish,
and we would obtain an incorrect result $\beta\equiv0$. It is well
known that the particle number is generically nonzero, except for
certain special cases when the particle production exactly vanishes. 

We conclude that the WKB ansatz approximates the mode function $v_{0}(t)$
insufficiently accurately for calculations of particle numbers. I
shall outline some known methods of improving the WKB approximation
in Sec.~\ref{sec:Perturbative-improvement-of}.

\subsection{Divergence of the WKB series\label{sub:Divergence-of-the}}

If the function $\omega(t)$ is analytic (or at least $C^{\infty}$)
in $t$, we may attempt to sum the series~(\ref{eq:W series}) by
evaluating the limit $n\rightarrow\infty$ at a fixed $t$. However,
the series~(\ref{eq:W series}) generally will not converge as $n\rightarrow\infty$.
Although the initial terms may decrease, eventually after a large
enough $n$ the terms $S_{n}$ will grow without bound. Thus the series~(\ref{eq:W series})
can be interpreted only as an asymptotic series, except for certain
special choices of $\omega(t)$. Below we shall obtain more precise
estimates of the growth of terms in that series, and presently we
demonstrate the generic divergence of the WKB series using qualitative
arguments.

Let us consider a frequency function $\omega(t)$ that allows an {}``in-out''
transition between $t=t_{0}$ and $t=t_{1}$, with super-adiabatic
regimes at $t\leq t_{0}$ and $t\geq t_{1}$. Suppose that the series~(\ref{eq:W series})
converges for some $\varepsilon=\varepsilon_{0}$ and for all $t$
within the relevant range $\left[t_{0},t_{1}\right]$. Then this series
will converge absolutely for smaller $\left|\varepsilon\right|<\varepsilon_{0}$,
yielding a well-defined function \begin{equation}
W_{\infty}(t;\varepsilon)\equiv\sum_{k=0}^{\infty}\varepsilon^{2k}S_{k}(t).\label{eq:W inf}\end{equation}
 By construction, the function $W_{\infty}(t;\varepsilon)$ is analytic
in $\varepsilon$ at $\varepsilon=0$. We shall now show that $W_{\infty}$
is an exact solution of Eq.~(\ref{eq:W equ}). Since the partial
sums $W_{n}$ were obtained by expanding the recurrence relation~(\ref{eq:W recur})
to a finite order in $\varepsilon$, we may substitute $W_{\infty}$
instead of $W_{n}$ in the r.h.s.~of Eq.~(\ref{eq:W recur}) and
find that\begin{equation}
W_{n}+O(\varepsilon^{2n+2})=\sqrt{\omega^{2}-\varepsilon^{2}\left[\frac{1}{2}\frac{W_{\infty}^{\prime\prime}}{W_{\infty}}-\frac{3}{4}\frac{W_{\infty}^{\prime2}}{W_{\infty}^{2}}\right]}\end{equation}
holds for all $n$. Therefore, the relation \begin{equation}
W_{\infty}=\sqrt{\omega^{2}-\varepsilon^{2}\left[\frac{1}{2}\frac{W_{\infty}^{\prime\prime}}{W_{\infty}}-\frac{3}{4}\frac{W_{\infty}^{\prime2}}{W_{\infty}^{2}}\right]}\label{eq:W eps equ}\end{equation}
holds as an identity between power series in $\varepsilon$ after
both sides are fully expanded. However, the r.h.s.~of Eq.~(\ref{eq:W eps equ})
can be also viewed as a power series in $\varepsilon$ in which only
the square root has been expanded, namely\begin{equation}
W_{\infty}=\omega-\frac{1}{2\omega}\varepsilon^{2}\left[\frac{1}{2}\frac{W_{\infty}^{\prime\prime}}{W_{\infty}}-\frac{3}{4}\frac{W_{\infty}^{\prime2}}{W_{\infty}^{2}}\right]+...\label{eq:W eps 2}\end{equation}
 Since the l.h.s.~of Eq.~(\ref{eq:W eps equ}) is an absolutely
convergent series, and the terms of such a series can be reordered
to form the r.h.s.~of Eq.~(\ref{eq:W eps 2}), both series specify
the same analytic function of $\varepsilon$. Hence, the series on
the r.h.s.~of Eq.~(\ref{eq:W eps 2}) converges and thus the function
$W_{\infty}(t;\varepsilon)$ is an exact solution of Eq.~(\ref{eq:W eps equ}). 

It follows that the {}``infinite-order'' WKB mode function\begin{equation}
v_{\infty}(t)=\frac{\sqrt{\varepsilon}}{\sqrt{W_{\infty}(t)}}\exp\left(-\frac{i}{\varepsilon}\int^{t}W_{\infty}(t')dt'\right)\end{equation}
 is an exact solution of Eq.~(\ref{eq:osc equ}). Then, from the
definition of the adiabatic vacuum state, we see that the mode function
$v_{\infty}(t)$ represents the adiabatic vacuum of \emph{every} order
and for \emph{all} fiducial times $t$ at once. In this uniquely identified
adiabatic vacuum state, the particle number is exactly zero at all
times and thus there is no (apparent) particle production at any time.
This outcome contradicts the fact that particle production in vacuum
is present in the generic case. Therefore in general the {}``infinite-order''
WKB solution $v_{\infty}(t)$ cannot exist.%
\footnote{Of course, an exact solution $W_{\infty}(t)$ of Eq.~(\ref{eq:W eps equ})
exists, but it cannot be exactly represented by the WKB series~(\ref{eq:W series})
because such $W_{\infty}(t)$ must be quickly oscillating and cannot
be a slow-changing function of $t$, as implicitly assumed by the
expansion~(\ref{eq:W series}). %
}

Another way to arrive at the same conclusion is to consider the particle
number $\left|\beta(t;\varepsilon)^{2}\right|$ in the second super-adiabatic
regime ($t\geq t_{1}$) as a function of $\varepsilon$. We can express
$\left|\beta(t;\varepsilon)\right|^{2}$ through the function $W_{\infty}(t;\varepsilon)$;
for example, using a zeroth-order adiabatic vacuum, we find\begin{equation}
\left|\beta(t;\varepsilon)\right|^{2}=\frac{\varepsilon}{4\omega W_{\infty}}\left[\frac{\left(\omega-W_{\infty}\right)^{2}}{\varepsilon^{2}}+\left(\frac{\omega'}{2\omega}-\frac{W_{\infty}^{\prime}}{2W_{\infty}}\right)^{2}\right].\end{equation}
 Note that the rapidly oscillating phase is absent from $\left|\beta\right|^{2}$
and that $W_{\infty}(t;\varepsilon=0)=\omega(t)$ for all $t$. Thus
$\left|\beta(t;\varepsilon)\right|^{2}$ is nonsingular (and equal
to zero) at $\varepsilon=0$. On the other hand, the particle number
is known%
\footnote{Kulsrud~\cite{Kul57} has shown that the change in the adiabatic
invariant is of order $\varepsilon^{p+2}$ if $\omega(t)$ has $p$
continuous derivatives. One can also demonstrate an exponential-like
behavior $\left|\beta\right|\propto\exp(-C/\varepsilon)$ for analytic
$\omega(t)$, under some technical conditions~\cite{JoyPfi91}. See
also Ref.~\cite{Ray80} for a derivation of an \emph{exact} invariant
which coincides with the adiabatic invariant in the leading order
in $\varepsilon$.%
} to decay faster than any power of $\varepsilon$ if the function
$\omega(t)$ is analytic or smooth (of class $C^{\infty}$). Thus
the particle number in the second super-adiabatic regime, $\left|\beta(t_{1};\varepsilon)\right|^{2}$,
is an analytic function of $\varepsilon$ which decays faster than
any power of $\varepsilon$ at $\varepsilon=0$. Such a function must
be identically equal to zero. It follows that the particle number
$\left|\beta(t_{1};\varepsilon)\right|^{2}$ would be identically
zero for all $\left|\varepsilon\right|<\varepsilon_{\max}$, given
the convergence assumption~(\ref{eq:W inf}). We know, however, that
generic choices of $\omega(t)$ involving an {}``in-out'' transition
will exhibit \emph{nonvanishing} particle production%
\footnote{Exceptions are provided by cases where all terms $S_{n}(t)$ of the
WKB series vanish after some $n$ (see e.g.~\cite{FroFro65}, chapter
2, for an explicit example), and also by functions $\omega(t)$ corresponding
to Schrödinger equations with exactly zero scattering~\cite{RobSal97}.%
} even for very small $\varepsilon$. We conclude that the WKB series
cannot converge for all $t$ within the interval $\left[t_{0},t_{1}\right]$,
except for special choices of $\omega(t)$ where the particle production
is absent. (Note that the WKB series \emph{converges} to $\omega(t)$
for $t$ within a super-adiabatic regime since all derivatives of
$\omega(t)$ vanish there, but nevertheless it does not yield a solution
$x(t)$ valid for all other $t$. We shall later show that in general
the WKB series cannot converge even for some special values of $t$
outside of super-adiabatic regimes.)

Finally, let us illustrate the growth of terms $S_{k}$ in the series~(\ref{eq:W series})
by performing an estimate of only the highest-order time derivatives
of $\omega(t)$ entering the expressions $S_{k}$. This estimate is
merely qualitative because in fact the contributions of lower-order
derivatives cannot be neglected; we shall obtain more precise results
in Sec.~\ref{sec:Properties-of-the}.

Rewriting the recurrence relation~(\ref{eq:W recur}) as\begin{equation}
W_{n+1}=\sqrt{\omega^{2}+\varepsilon^{2}\sqrt{W_{n}}\left(\frac{1}{\sqrt{W_{n}}}\right)^{\prime\prime}},\end{equation}
we find after some algebra that the term $S_{n}$ contains the highest-order
derivative of $\omega(t)$ always in the following combination, \begin{equation}
S_{n}(t)=\frac{\left(-1\right)^{n}\varepsilon^{2n}}{4^{n}\omega^{2n}}\frac{d^{2n}}{dt^{2n}}\frac{1}{\sqrt{\omega(t)}}+(\textrm{lower-order}),\label{eq:Sk approx}\end{equation}
where we have suppressed terms containing lower-order derivatives
of $\omega$. Now we need to estimate the growth of derivatives of
$1/\sqrt{\omega}$. For a generic analytic function $f(z)$ having
some poles in the complex $z$ plane, the growth of derivatives at
$z=z_{0}$ can be estimated as\begin{equation}
\left.\frac{d^{n}f(z)}{dz^{n}}\right|_{z=z_{0}}\sim\frac{n!}{\left(z_{1}-z_{0}\right)^{n+1}},\end{equation}
where $z_{1}$ is the pole of $f(z)$ nearest to $z=z_{0}$ (see Appendix~\ref{sec:Growth-of-derivatives}
for a derivation of this formula). In our case, poles of $1/\sqrt{\omega}$
correspond to zeros of $\omega(t)$, i.e.~turning points in the complex
$t$ plane. Let us suppose that the nearest zero of $\omega(t)$ is
at $t=t_{1}$. Under the assumption that the value of $S_{n}$ is
of order of the first term in Eq.~(\ref{eq:Sk approx}), we find\begin{equation}
\left|S_{n}(t)\right|\sim\left(\frac{\varepsilon}{2\omega(t)}\right)^{2n}\frac{\left(2n\right)!}{\left|t-t_{1}\right|^{2n+1}}.\end{equation}
It is now straightforward to see that the terms $S_{n}$ may decrease
for small $n$ but eventually begin to grow after $n_{*}\sim\varepsilon^{-1}\omega(t)\left|t-t_{1}\right|$.
The term $S_{n_{*}}$ is of order\begin{equation}
\left(\frac{\varepsilon}{2\omega(t)}\right)^{2n_{*}}\frac{\left(2n_{*}\right)!}{\left|t-t_{1}\right|^{2n_{*}+1}}\sim\frac{1}{\sqrt{n_{*}}}\exp\left(-2n_{*}\right).\end{equation}
 The above values are roughly correct order-of-magnitude estimates
for the optimal number of terms in the WKB series and for the optimal
precision, as we show below.

\section{Perturbative improvement of WKB\label{sec:Perturbative-improvement-of}}

The precision of the WKB approximation can be improved by applying
a time-dependent perturbation theory to the WKB ansatz. Equivalent
and closely related perturbative methods were used e.g.~in~\cite{ZelSta71,GriMamMos94}
for particle production calculations and in the mathematical literature~\cite{Hea62,Olv74}
for precision estimates of the WKB ansatz. A related technique called
\textsl{\emph{{}``quasilinearization}}''~\cite{BelKal65} was applied
in~e.g.~\cite{DatRam81,KriMan04} to obtain improved approximations
to bound states of the Schrödinger equation (although the convergence
of the method was investigated only numerically). I shall now give
a brief overview of these methods.

\subsection{Methods based on the Riccati equation }

The simplest version of the perturbative improvement technique is
intended to determine a small correction to an approximate positive-frequency
solution of Eq.~(\ref{eq:osc equ T}),\begin{equation}
x_{+}(t)=C\exp\left[-i\int_{0}^{t}\omega(t')dt'\right],\end{equation}
in the form\begin{equation}
x(t)=x_{+}(t)\exp\left[\int_{0}^{t}y(t')dt'\right].\end{equation}
 Thus one introduces a new (complex-valued) dependent variable $y(t)$
via\begin{equation}
y(t)\equiv\frac{\dot{x}}{x}+i\omega(t).\label{eq:x and y}\end{equation}
The equation for $y(t)$ is easily derived and is a Riccati equation,\begin{equation}
\dot{y}-2i\omega y=i\dot{\omega}-y^{2}.\label{eq:y equ}\end{equation}
By assumption, $x(t)$ differs little from $x_{+}(t)$, and so one
expects that the value of $y(t)$ is small and that the $y^{2}$ term
in Eq.~(\ref{eq:y equ}) can be treated as a perturbation. Disregarding
the $y^{2}$ term and using the natural initial condition $y(0)=0$,
one finds the approximate solution\begin{equation}
y_{(1)}(t)=i\int_{0}^{t}\dot{\omega}(t')\exp\left[2i\int_{t'}^{t}\omega(t^{\prime\prime})dt^{\prime\prime}\right]dt^{\prime}.\label{eq:y1}\end{equation}
This first approximation $y_{1}(t)$ is already sufficient to compute
the leading contribution to the exponentially small particle occupation
number in the context of an {}``in-out'' transition. The resulting
(approximate) value of $\beta$ is found after some algebra as\begin{equation}
\beta_{(1)}(t_{1})=\frac{v_{0}(t_{1})}{2i\sqrt{\omega_{1}}}y_{(1)}(t_{1}).\end{equation}
 For analytic functions $\omega(t)$, the value of $\beta_{(1)}(t_{1})$
is typically exponentially small due to rapid oscillations of the
integrand in Eq.~(\ref{eq:y1}).

A variation of this perturbative improvement technique was developed
and used in Ref.~\cite{HonVilWin03}. There the new dependent variable
for Eq.~(\ref{eq:osc equ T}) was defined by\begin{equation}
\zeta(t)\equiv\frac{\dot{x}+i\omega x}{\dot{x}-i\omega x}.\end{equation}
The function $\zeta(t)$ satisfies the equation\begin{equation}
\dot{\zeta}-2i\omega\zeta=-\frac{\dot{\omega}}{2\omega}\left(1-\zeta^{2}\right)\label{eq:zeta equ}\end{equation}
and can be interpreted as the instantaneous squeezing parameter describing
the squeezed state of one mode of the quantum field at time $t$ relative
to the instantaneous vacuum state at that time. The particle number
is found from\begin{equation}
\left|\beta(t)\right|^{2}=\frac{\left|\zeta(t)\right|^{2}}{1-\left|\zeta(t)\right|^{2}}.\label{eq:beta thru zeta}\end{equation}
 The leading-order solution $\zeta_{(1)}$ of Eq.~(\ref{eq:zeta equ})
with the initial condition $\zeta(0)=0$ is obtained by disregarding
$\zeta^{2}$, \begin{equation}
\zeta_{(1)}(t)=-\int_{0}^{t}\frac{\dot{\omega}(t')dt'}{2\omega(t')}\exp\left[2i\int_{t'}^{t}\omega(t^{\prime\prime})dt^{\prime\prime}\right].\label{eq:zeta 1}\end{equation}
Further approximations $\zeta_{(n)}(t)$ are found from the recurrence
relation\begin{equation}
\zeta_{(n+1)}(t)=-\!\int_{0}^{t}\frac{\dot{\omega}dt'}{2\omega}\!\left[1-\zeta_{(n)}^{2}(t')\right]\exp\left[2i\!\int_{t'}^{t}\!\omega(t^{\prime\prime})dt^{\prime\prime}\right].\end{equation}
It was shown in Ref.~\cite{HonVilWin03} that the sequence $\zeta_{(1)}$,
$\zeta_{(2)},...$ converges to the solution of Eq.~(\ref{eq:zeta equ})
as long as $\left|\zeta_{(n)}(t)\right|<1$. An advantage of using
the variable $\zeta(t)$ instead of $y(t)$ is that the values of
$\zeta$ are always bounded as long as $\omega(t)\neq0$, which helps
in the analysis. (It is straightforward to verify that Eq.~(\ref{eq:vk norm})
together with $\textrm{Im }\omega=0,\,\omega\neq0$ yield the bound
$\left|\zeta\right|<1$.) 

The method of {}``quasilinearization''~\cite{BelKal65} uses the
ansatz\begin{equation}
x_{(2)}(t)=C\exp\left[\int_{0}^{t}\left(y_{(2)}(t)+y_{(1)}(t)-i\omega\right)dt'\right],\end{equation}
which leads to the easily solvable equation\begin{equation}
\dot{y}_{(2)}-2\left(i\omega-y_{(1)}\right)y_{(2)}=-y_{(1)}^{2},\end{equation}
where the term quadratic in $y_{(2)}$ has been neglected and $y_{(1)}$
is given by Eq.~(\ref{eq:y1}). Further corrections $y_{(3)},y_{(4)},...$
and the corresponding solutions $x_{(3)},x_{(4)},...$ are determined
in the same manner. Analytic and numerical studies~\cite{KriMan04}
show that the accuracy of the solution $x_{(n)}(t)$ improves quadratically
with $n$, the error being of order $\varepsilon^{2^{n}}$. However,
precise conditions for the convergence of this method do not seem
to have been investigated. 

The methods outlined so far are conceptually simple but lead to nonlinear
equations which complicates their analysis. We shall therefore use
an equivalent but somewhat more long-winded perturbative technique
based on a system of two linear equations. The advantage will be that
we shall obtain approximations to the solution $x(t)$ in the form
of a series (called the Bremmer series).

\subsection{The Bremmer series \label{sub:The-Bremmer-series}}

In this section I mostly follow Refs.~\cite{Atk60,Kay61}. The usual
WKB approximation to solutions of Eq.~(\ref{eq:osc equ}) is \begin{equation}
x(t)\approx AX_{+}(t)+BX_{-}(t),\end{equation}
where $A,B$ are constants and\begin{equation}
X_{\pm}(t)\equiv\frac{1}{\sqrt{\omega(t)}}\exp\left[\mp\frac{i}{\varepsilon}\int_{t_{0}}^{t}\omega(t')dt'\right]\label{eq:Xpm def}\end{equation}
are the two WKB branches (here, $t_{0}$ is an arbitrary initial point).
To improve this approximation, one looks for solutions $x(t)$ in
the form\begin{equation}
x(t)=p(t)X_{+}(t)+q(t)X_{-}(t),\label{eq:Bremmer ansatz}\end{equation}
where the coefficients $p(t),q(t)$ are now time-dependent. By introducing
two unknown functions $p(t),q(t)$ instead of one unknown $x(t)$,
we have added a degree of freedom which is canceled by imposing the
additional relation\begin{equation}
\dot{x}(t)=\frac{i}{\varepsilon}\omega(t)\left[-p(t)X_{+}(t)+q(t)X_{-}(t)\right].\end{equation}
This relation signifies that the derivative of $x(t)$ can be computed
from Eqs.~(\ref{eq:Xpm def})-(\ref{eq:Bremmer ansatz}) by formally
treating $p(t)$, $q(t)$, and $\omega(t)$ as constants. It is then
straightforward to derive the following simple equations for $p(t)$
and $q(t)$,\begin{equation}
\dot{p}=\frac{1}{2}\frac{\dot{\omega}}{\omega}\frac{X_{-}}{X_{+}}q,\quad\dot{q}=\frac{1}{2}\frac{\dot{\omega}}{\omega}\frac{X_{+}}{X_{-}}p.\label{eq:p q equ}\end{equation}

In the context of QFTCS, one is usually looking for small corrections
to the positive-frequency WKB solution $x(t)=X_{+}(t)$. Therefore
we assume the initial conditions $p(t_{0})=1$, $q(t_{0})=0$, and
rewrite Eqs.~(\ref{eq:p q equ}) as the following integral equations,\begin{equation}
p(t)=1+\frac{1}{2}\int_{t_{0}}^{t}\!\frac{\dot{\omega}}{\omega}\frac{X_{-}}{X_{+}}qdt,\quad q(t)=\frac{1}{2}\int_{t_{0}}^{t}\!\frac{\dot{\omega}}{\omega}\frac{X_{+}}{X_{-}}pdt.\end{equation}
These equations can be solved iteratively, starting from the initial
approximation $p(t)\equiv1$, $q(t)\equiv0$. The result is conveniently
written as a series involving an auxiliary sequence $u_{n}(t)$,\begin{align}
p(t) & =1+\sum_{n=1}^{\infty}u_{2n}(t),\quad q(t)=\sum_{n=1}^{\infty}u_{2n-1}(t),\end{align}
where the functions $u_{n}(t)$, $n\geq0$, are defined recursively
starting from $u_{0}(t)\equiv1$ as\begin{align}
u_{2n}(t) & =\frac{1}{2}\int_{t_{0}}^{t}\frac{\dot{\omega}}{\omega}\frac{X_{-}}{X_{+}}u_{2n-1}dt,\quad n\geq1,\label{eq:u2n}\\
u_{2n+1}(t) & =\frac{1}{2}\int_{t_{0}}^{t}\frac{\dot{\omega}}{\omega}\frac{X_{+}}{X_{-}}u_{2n}dt,\quad n\geq0.\label{eq:u2np1}\end{align}
The resulting infinite series for the solution $x(t)$, \begin{equation}
x(t)=X_{+}+\sum_{n=1}^{\infty}\left(u_{2n-1}X_{-}+u_{2n}X_{+}\right),\label{eq:Bremmer series def}\end{equation}
is called the \textsl{Bremmer series}. A physical motivation behind
this derivation is given in~\cite{Lan51} and references therein.

\subsection{Convergence of the Bremmer series}

The Bremmer series converges to the exact solution $x(t)$ absolutely
and uniformly for $t_{0}<t<t_{\max}$ (where $t_{\max}$ may be infinite)
under the rather weak restriction%
\footnote{The condition~(\ref{eq:Bremmer conv cond}) would be violated, for
instance, in the presence of parametric resonance with $t_{\max}=\infty$.%
} \begin{equation}
\int_{t_{0}}^{t_{\max}}\left|\frac{\dot{\omega}}{\omega}\right|dt<\infty.\label{eq:Bremmer conv cond}\end{equation}
 The convergence can be demonstrated by the following argument~\cite{Atk60,Kay61,KelKel62}.
The sequence $u_{n}(t)$ is majorized by the auxiliary sequence $U_{n}(t)$
defined by\begin{equation}
U_{n+1}(t)=\int_{t_{0}}^{t}\left|\frac{\dot{\omega}}{2\omega}\right|U_{n}dt,\quad U_{0}(t)\equiv1;\end{equation}
namely $\left|u_{n}(t)\right|\leq U_{n}(t)$ for all $n,t$. Convergence
of the majorizing series $\sum_{n=0}^{\infty}U_{n}(t)$ is therefore
sufficient for the absolute convergence of the Bremmer series. In
turn, the series $\sum_{n=0}^{\infty}U_{n}(t)\equiv U(t)$ can be
summed explicitly by deriving a differential equation for $U(t)$,\begin{equation}
\dot{U}(t)=\left|\frac{\dot{\omega}}{2\omega}\right|U(t),\quad U(t_{0})=1.\end{equation}
The solution is\begin{equation}
U(t)=\exp\left[\int_{t_{0}}^{t}\left|\frac{\dot{\omega}}{2\omega}\right|dt\right].\label{eq:U sol}\end{equation}
Thus the Bremmer series converges as long as the integral in Eq.~(\ref{eq:U sol})
is finite, which yields the condition~(\ref{eq:Bremmer conv cond}).
The convergence is uniform in $t$ because an upper bound\begin{equation}
\int_{t_{0}}^{t_{\max}}\left|\frac{\dot{\omega}}{2\omega}\right|dt<M\end{equation}
entails\begin{equation}
\left|u_{n}(t)\right|\leq U_{n}(t)<\frac{1}{n!}M^{n},\quad t_{0}<t<t_{\max},\end{equation}
and so the number of terms $n$ may be chosen in advance to guarantee
a desired precision for all $t<t_{\max}$. Namely, it is straightforwardly
seen that a relative precision $e^{-P}$ will be achieved by computing
$n=O(P/\ln P)$ terms.

\section{Precision of the WKB approximation\label{sec:Properties-of-the}}

\subsection{The adiabatic expansion}

The WKB series~(\ref{eq:W series}) entails an adiabatic expansion
of the WKB solution\begin{equation}
x_{\textrm{WKB}}(t)=\frac{1}{\sqrt{W_{n}(t)}}\exp\left[-\frac{i}{\varepsilon}\int_{t_{0}}^{t}W_{n}(t')dt'\right]\end{equation}
 in powers of the adiabatic parameter $\varepsilon$,\begin{align}
x_{\textrm{WKB}}(t)=X_{+}(t) & \left[1+\frac{i\varepsilon}{4}\int_{t_{0}}^{t}\left(\frac{\ddot{\omega}}{\omega^{2}}-\frac{3}{2}\frac{\dot{\omega}^{2}}{\omega^{3}}\right)dt'\right.\nonumber \\
 & \,\left.+\frac{\varepsilon^{2}}{8}\left(\frac{\ddot{\omega}}{\omega^{2}}-\frac{3}{2}\frac{\dot{\omega}^{2}}{\omega^{3}}\right)+...\right],\label{eq:xWKB 1}\end{align}
where $X_{+}(t)$ is defined by Eq.~(\ref{eq:Xpm def}). The series
in brackets above contains terms produced by expanding the denominator
$1/\sqrt{W_{n}}$ as well as by expanding the exponential. Therefore
we expect that the series in Eq.~(\ref{eq:xWKB 1}) is again a divergent
asymptotic series. We shall now analyze this series to determine the
optimal number of terms and the best attainable precision. 

The main idea of this analysis is to represent the exact solution
$x(t)$ by the (convergent) Bremmer series and to compare the latter
with the WKB series~(\ref{eq:xWKB 1}). However, the structure of
the Bremmer series differs from that of the WKB series in two aspects.
Firstly, the Bremmer series involves both positive- and negative-frequency
branches while the WKB ansatz~(\ref{eq:xWKB 1}) contains only the
positive-frequency branch. Secondly, the Bremmer series is not a power
series in $\varepsilon$, but instead it involves oscillating integrals
containing $\varepsilon^{-1}$ under exponentials. Thus we need to
obtain an asymptotic expansion of the Bremmer series ansatz~(\ref{eq:Bremmer ansatz})
in powers of $\varepsilon$, of the form\begin{equation}
p(t)X_{+}+q(t)X_{-}=X_{+}\sum_{k=0}^{\infty}A_{k}\varepsilon^{k}+X_{-}\sum_{k=0}^{\infty}B_{k}\varepsilon^{k},\label{eq:Bremmer exp 1}\end{equation}
where $A_{k}$ and $B_{k}$ are some time-dependent coefficients.
The positive-frequency part ($X_{+}\sum_{k}A_{k}\varepsilon^{k}$)
of the above expansion will coincide with the expansion~(\ref{eq:xWKB 1})
because the asymptotic expansion in power series is unique~\cite{Din73}.
Note that the split between positive- and negative-frequency branches
in the expansion in the r.h.s.~of Eq.~(\ref{eq:Bremmer exp 1})
is well-defined~\cite{CosDupKru04}, but nevertheless $\sum_{k}A_{k}\varepsilon^{k}$
is \emph{not} an expansion of $p(t)$, and neither is $\sum_{k}B_{k}\varepsilon^{k}$
an expansion of $q(t)$. We shall now derive an asymptotic expansion
of the form~(\ref{eq:Bremmer exp 1}) from Eqs.~(\ref{eq:u2n})-(\ref{eq:u2np1}).
It will turn out that only the positive-frequency branch survives
in the asymptotic expansion, i.e.~$B_{k}\equiv0$, which makes the
correspondence with the WKB series immediate.

Let us pass from the time variable $t$ to the dimensionless {}``phase''
variable $\theta$, \begin{equation}
t\rightarrow\theta(t)\equiv\int_{t_{0}}^{t}\omega(t')dt'.\label{eq:theta def}\end{equation}
This is a well-defined transformation as long as $\omega(t)$ remains
real and nonzero in the relevant range of $t$, and in that case $\omega(t)$
remains an analytic function of $\theta$. Equations~(\ref{eq:u2n})-(\ref{eq:u2np1})
are then rewritten as\begin{align}
u_{2n}(\theta) & =\int_{0}^{\theta}e^{2i\theta'/\varepsilon}\alpha(\theta')u_{2n-1}(\theta')d\theta',\quad n\geq1,\label{eq:u2n theta}\\
u_{2n+1}(\theta) & =\int_{0}^{\theta}e^{-2i\theta'/\varepsilon}\alpha(\theta')u_{2n}(\theta')d\theta',\quad n\geq0,\label{eq:u2n1 theta}\end{align}
where for convenience we have introduced the dimensionless function\begin{equation}
\alpha(\theta)\equiv\left.\frac{1}{2\omega^{2}}\frac{d\omega}{dt}\right|_{t\rightarrow\theta}.\label{eq:alpha def}\end{equation}
Note that $\varepsilon\alpha(\theta)\ll1$ under the adiabaticity
condition~(\ref{eq:adiab cond}).

Starting from Eqs.~(\ref{eq:u2n theta})-(\ref{eq:u2n1 theta}) with
$u_{0}(\theta)\equiv1$, we shall first obtain expansions for $u_{1}(\theta)$
and $u_{2}(\theta)$, which will make further calculations more transparent.
To expand $u_{1}(\theta)$ in powers of $\varepsilon$, we repeatedly
apply integration by parts to Eq.~(\ref{eq:u2n1 theta}) with $n=0$
and find\begin{equation}
u_{1}(\theta)=-\left.\left[e^{-2i\theta'/\epsilon}\left(\frac{\varepsilon\alpha}{2i}+\frac{\varepsilon^{2}\alpha^{\prime}}{\left(2i\right)^{2}}+\frac{\varepsilon^{3}\alpha^{\prime\prime}}{\left(2i\right)^{3}}+...\right)\right]\right|_{\theta'=0}^{\theta'=\theta},\label{eq:u1 pre asympt}\end{equation}
 where the prime denotes derivatives with respect to $\theta$. We
further assume that the point $\theta=0$ ($t=t_{0}$) is located
within a super-adiabatic regime where $\alpha(\theta)$ and all its
derivatives vanish. Then Eq.~(\ref{eq:u1 pre asympt}) simplifies
to\begin{equation}
u_{1}(\theta)=-\frac{\varepsilon}{2i}e^{-2i\theta/\varepsilon}\sum_{k=0}^{\infty}\frac{\varepsilon^{k}}{\left(2i\right)^{k}}\frac{d^{k}\alpha(\theta)}{d\theta^{k}}.\label{eq:u1 asympt}\end{equation}

The expansion for $u_{2}(\theta)$ is found from Eq.~(\ref{eq:u2n theta})
as\begin{equation}
u_{2}(\theta)=-\frac{\varepsilon}{2i}\sum_{k=0}^{\infty}\frac{\varepsilon^{k}}{\left(2i\right)^{k}}\int_{0}^{\theta}\alpha(\theta')\frac{d^{k}\alpha(\theta')}{d\theta^{\prime k}}d\theta';\label{eq:u2 asympt}\end{equation}
note the absence of the quickly oscillating exponential. Hence, the
initial terms of the expansion~(\ref{eq:Bremmer exp 1}) are \begin{equation}
X_{+}(1+u_{2})+X_{-}u_{1}=X_{+}\!\left[1+\frac{i\varepsilon}{4}{\textstyle \int}\big(\frac{\ddot{\omega}}{\omega^{3}}-\frac{3}{2}\frac{\dot{\omega}^{2}}{\omega^{4}}\big)dt+...\right],\end{equation}
which reproduces the initial terms of Eq.~(\ref{eq:xWKB 1}). The
negative-frequency branch is absent due to the factor $\exp(-2i\theta/\varepsilon)$
in $u_{1}$. It is easy to see that further terms $u_{n}$ of the
Bremmer series have expansions of the same form: namely, the odd-numbered
terms $u_{2n+1}$ contain the oscillating factor while the even-numbered
terms $u_{2n}$ do not. For convenience, we separate these factors
explicitly and define the auxiliary functions $\lambda_{n}(\theta)$
and $\mu_{n}(\theta)$ by\begin{align}
u_{2n-1}(\theta) & \equiv\left(-\frac{\varepsilon}{2i}\right)^{n}e^{-2i\theta/\varepsilon}\lambda_{n}(\theta),\quad n\geq1;\\
u_{2n}(\theta) & \equiv\left(-\frac{\varepsilon}{2i}\right)^{n}\mu_{n}(\theta),\quad n\geq0.\end{align}
The functions $\lambda_{n},\mu_{n}$ are (formal) series in $\varepsilon$
that satisfy the following recurrence relations, \begin{align}
\lambda_{n+1}(\theta) & =\sum_{k=0}^{\infty}\left(\frac{\varepsilon}{2i}\right)^{k}\frac{d^{k}}{d\theta^{k}}\left[\alpha(\theta)\mu_{n}(\theta)\right],\quad n\geq0;\label{eq:lambda n1 rel}\\
\mu_{n}(\theta) & =\int_{0}^{\theta}\alpha(\theta')\lambda_{n}(\theta')d\theta',\quad n\geq1;\quad\mu_{0}\equiv1.\label{eq:mu n rel}\end{align}
In terms of these functions, the asymptotic expansion of the Bremmer
series can be expressed as\begin{align}
p(\theta)X_{+}(\theta)+q(\theta)X_{-}(\theta) & \equiv X_{+}(\theta)\sum_{k=0}^{\infty}A_{k}\varepsilon^{k}\nonumber \\
 & =X_{+}(\theta)\sum_{k=0}^{\infty}\left(-\frac{\varepsilon}{2i}\right)^{k}\left(\lambda_{k}+\mu_{k}\right),\label{eq:Bremmer exp 2}\end{align}
where we have formally defined $\lambda_{0}\equiv0$. Since the r.h.s.~of
Eq.~(\ref{eq:Bremmer exp 2}) will coincide with the WKB series~(\ref{eq:xWKB 1})
after $\lambda_{k}$ and $\mu_{k}$ are fully expanded in powers of
$\varepsilon$, it is sufficient to analyze the convergence of these
latter expansions. Note that the $n$-th order WKB approximation corresponds
to retaining the terms of the series~(\ref{eq:Bremmer exp 2}) up
to $k=2n$.

\subsection{Precision of the asymptotic expansion}

The goal of this section is to derive the optimal truncation of the
asymptotic series~(\ref{eq:Bremmer exp 2}). We begin by analyzing
just the first two terms of that expansion, namely\begin{equation}
(u_{0}+u_{2})X_{+}+u_{1}X_{-}=X_{+}\left[1-\frac{\varepsilon}{2i}\left(\lambda_{1}+\mu_{1}\right)\right].\end{equation}
 Subsequently we shall examine the convergence of the terms $\lambda_{k},\mu_{k}$
with $k>1$.

We find from Eqs.~(\ref{eq:u1 asympt}) and (\ref{eq:u2 asympt})
that the quantities $\lambda_{1}$ and $\mu_{1}$ are expressed by
the series\begin{align}
\lambda_{1}(\theta) & =\sum_{k=0}^{\infty}\frac{\varepsilon^{k}}{\left(2i\right)^{k}}\frac{d^{k}\alpha(\theta)}{d\theta^{k}},\label{eq:lambda1 series}\\
\mu_{1}(\theta) & =\sum_{k=0}^{\infty}\frac{\varepsilon^{k}}{\left(2i\right)^{k}}\int_{0}^{\theta}\alpha(\theta')\frac{d^{k}\alpha(\theta')}{d\theta^{\prime k}}d\theta'.\end{align}
 These series are typically divergent because derivatives $d^{k}\alpha/d\theta^{k}$
of an analytic function $\alpha(\theta)$ grow as $k!$ with $k\rightarrow\infty$.
The growth of such derivatives is determined by the location of the
singularities of $\alpha(\theta)$ in the complex $\theta$ plane
(see Appendix~\ref{sec:Growth-of-derivatives} for more details).
For instance, if the singularity of $\alpha(\theta)$ nearest to the
point $\theta$ is a simple pole at $\theta_{1}$ with residue $-c_{1}$,
then for large $k$ there is an asymptotic estimate\begin{equation}
\frac{d^{k}\alpha(\theta)}{d\theta^{k}}\approx\frac{c_{1}k!}{\left(\theta_{1}-\theta\right)^{k+1}}.\label{eq:f growth estimate}\end{equation}
 In the present case, the function $\alpha(\theta)$ defined by Eq.~(\ref{eq:alpha def})
has simple poles at the locations $\theta_{1},\theta_{2},...$ corresponding
to \emph{zeros} $t_{1},t_{2},...$ of $\omega(t)$, namely\begin{equation}
\theta_{j}=\int_{t_{0}}^{t_{j}}\omega(t)dt,\quad j=1,2,...;\quad\omega(t_{j})=0.\label{eq:theta through t}\end{equation}
 A \emph{pole} of $\omega(t)$ corresponds to $\theta=\infty$ and
typically will not be the nearest singularity of $\alpha(\theta)$.
If $\omega(t)$ has a zero of $\nu$-th order at $t=t_{j}$, i.e.~$\omega(t)\sim(t-t_{j})^{\nu}$,
the corresponding residue of the simple pole of $\alpha(\theta)$
will be \begin{equation}
-c_{j}=\frac{1}{2}\frac{\nu}{\nu+1}.\end{equation}

Assuming that $\theta_{1}$ is the nearest pole of $\alpha(\theta)$,
it is straightforward to find that the terms of the series~(\ref{eq:u1 asympt})
will start growing after \begin{equation}
k>k_{*}(\theta;\varepsilon)\equiv2\varepsilon^{-1}\left|\theta_{1}-\theta\right|=2\varepsilon^{-1}\left|\int_{t}^{t_{1}}\omega(t')dt'\right|.\label{eq:kstar est}\end{equation}
If there is any ambiguity in the choice of the complex integration
contours in Eqs.~(\ref{eq:theta through t}) and (\ref{eq:kstar est}),
the contours should be such as to yield the smallest value of $\left|\theta_{1}-\theta\right|$
since we require $\theta_{1}$ to be the nearest singularity to $\theta$. 

The best attainable precision $\delta u_{1}$ is of order of the smallest
retained term, which can be estimated using Eq.~(\ref{eq:f growth estimate})
and Stirling's formula, $k!\approx\sqrt{2\pi k}e^{-k}k^{k}$, as\begin{equation}
\delta u_{1}=\left(\frac{\varepsilon}{2}\right)^{k_{*}+1}\left|\frac{d^{k_{*}}\alpha(\theta)}{d\theta^{k_{*}}}\right|\approx\left|c_{1}\right|\sqrt{\frac{2\pi}{k_{*}}}e^{-k_{*}}.\label{eq:error u1}\end{equation}

We shall show below that higher terms $u_{k}$, $k\geq3$, of the
Bremmer series have better convergence behavior and their error is
dominated by that given in Eq.~(\ref{eq:error u1}). Thus the formulae
(\ref{eq:kstar est})-(\ref{eq:error u1}) are the central result
of the present paper. The order $n$ of the WKB approximation, as
defined in Sec.~\ref{sub:The-WKB-approximation}, is related to the
order $k$ of the corresponding Bremmer series by \begin{equation}
n_{\max}=\frac{1}{2}k_{*}(t;\varepsilon)=\varepsilon^{-1}\left|\int_{t}^{t_{1}}\omega(t')dt'\right|.\label{eq:nmax k}\end{equation}
 Hence, the optimal order of the WKB approximation and the optimal
precision are given by Eqs.~(\ref{eq:nmax def}) and (\ref{eq:error nmax}),
where we have set $\varepsilon\equiv1$. The prefactor $c_{1}$ is
of order unity and will be relatively unimportant for our considerations.

Let us now analyze the convergence of the series~(\ref{eq:u2 asympt}).
After repeated integration by parts, we have \begin{equation}
\int\alpha\alpha^{(k)}d\theta=\alpha\alpha^{(k-1)}-\alpha^{\prime}\alpha^{(k-2)}+...,\end{equation}
and the first term dominates for large $k$. Thus we find the following
estimate for large $k$,\begin{equation}
\int_{0}^{\theta}\alpha\alpha^{(k)}d\theta'\approx\alpha(\theta)\alpha^{(k-1)}(\theta)\approx\alpha(\theta)\frac{c_{1}\left(k-1\right)!}{\left(\theta_{1}-\theta\right)^{k}}.\end{equation}
Hence, the asymptotic series~(\ref{eq:u2 asympt}) starts to diverge
after $k>1+k_{*}(\theta;\varepsilon)$, i.e.~one term later than
the series for $u_{1}$, and the best attainable precision $\delta u_{2}$
is\begin{equation}
\delta u_{2}=\left(\frac{\varepsilon}{2}\right)^{k_{*}+2}\alpha(\theta)\left|\frac{d^{k_{*}}\alpha(\theta)}{d\theta^{k_{*}}}\right|\approx\frac{\varepsilon\alpha(\theta)}{2}c_{1}\sqrt{\frac{2\pi}{k_{*}}}e^{-k_{*}}.\end{equation}
Note that $\delta u_{2}=\left(\varepsilon\alpha/2\right)\delta u_{1}$,
while $\varepsilon\alpha\ll1$ due to the adiabaticity condition,
hence $\delta u_{2}\ll\delta u_{1}$. We find that the precision of
the partial asymptotic series $\lambda_{1}+\mu_{1}$ is limited by
the error in $\lambda_{1}$, which is estimated by Eq.~(\ref{eq:error u1}). 

It remains to analyze the behavior of the higher-order terms in the
expansion~(\ref{eq:Bremmer exp 2}). It follows from Eq.~(\ref{eq:lambda n1 rel})
that the series for $\lambda_{n}$ is always a sum of several series
of the form\begin{equation}
\sum_{k=0}^{\infty}\left(\frac{\varepsilon}{2i}\right)^{k}\frac{d^{k}}{d\theta^{k}}f(\theta),\label{eq:lambda n series}\end{equation}
where $f(\theta)$ is some analytic function expressed through $\alpha(\theta)$
and its derivatives. The functions $f(\theta)$ will have singularities
at the same points in the complex $\theta$ plane as the function
$\alpha(\theta)$. Thus each series of the form~(\ref{eq:lambda n series})
will have essentially the same convergence properties as the series
for $\lambda_{1}(\theta)$ analyzed above; namely, the series will
start diverging after $k_{*}$ terms and have an exponentially small
error term. However, in the expansion~(\ref{eq:Bremmer exp 2}) the
series $\lambda_{n}$ is multiplied by $\varepsilon^{n}$. Thus, the
series for $u_{2n-1}$, where $n>1$, always converges better (i.e.~it
starts to diverge at a later term) than the series~(\ref{eq:u1 asympt})
for $u_{1}$, and the lowest error in $u_{2n-1}$ is always smaller
than that of $u_{1}$. The same considerations can be seen to apply
to the series for $\mu_{n}$, $n>1$, and hence for the terms $u_{2n}$,
$n>1$, of the Bremmer series. Therefore, the optimal truncation of
the expansion~(\ref{eq:Bremmer exp 2}) is the same as the optimal
truncation of the series $\lambda_{1}(\theta)$, as given by Eq.~(\ref{eq:kstar est}),
and the lowest error is dominated by the error in $\lambda_{1}$,
as estimated by Eq.~(\ref{eq:error u1}). This argument concludes
the derivation of Eqs.~(\ref{eq:nmax def})-(\ref{eq:error nmax}).

\subsection{Particle production and the definition of vacuum \label{sub:Particle-production-and}}

We have seen that the WKB series diverges unless there is exactly
no particle production. Furthermore, we shall now show that the ambiguity
in the definition of vacuum is quantitatively related to the presence
of particle production. 

Let us consider an {}``in-out'' transition between two super-adiabatic
epochs and assume that there is nonzero particle creation, as there
will be in the generic case. Intuitively one expects that most of
the particles are created during the epoch where gravity is most significant
(the {}``active'' epoch). However, the adiabatic vacuum during the
active epoch is defined only up to the precision of the WKB approximation
which is fundamentally limited. The precision is improved as we move
away from the active epoch, but so does the expected rate of particle
production. We shall now derive some estimates that illustrate this
connection.

As we have seen, the optimal precision of the WKB approximation is
of the order $\exp(-2n_{\max})$, where $n_{\max}$ depends on the
time at which the WKB approximation is applied. We shall use the variable
$\theta$ defined by Eq.~(\ref{eq:theta def}) as the time variable.
If the active epoch corresponds to values around $\theta=\theta_{0}$,
then, according to Eq.~(\ref{eq:nmax k}),\begin{equation}
n_{\max}(\theta_{0})=\epsilon^{-1}\left|\theta_{0}-\theta_{1}\right|,\end{equation}
where $\theta_{1}$ is the pole of $\alpha(\theta)$ closest to $\theta=\theta_{0}$
in the complex $\theta$ plane, and the function $\alpha(\theta)$
is defined by Eq.~(\ref{eq:alpha def}). Now consider the total particle
number produced during the active epoch. As reviewed in Sec.~\ref{sec:Perturbative-improvement-of},
the leading contribution to this particle number is exponentially
small in the adiabatic parameter and can be computed from a perturbatively
improved WKB approximation. For instance, using Eqs.~(\ref{eq:beta thru zeta})
and (\ref{eq:zeta 1}) yields\begin{equation}
\beta\sim\int_{-\infty}^{+\infty}e^{-2i\theta/\varepsilon}\alpha(\theta)d\theta.\label{eq:beta integral}\end{equation}
Since by assumption the function $\alpha(\theta)$ is analytic, the
integral involved in Eq.~(\ref{eq:beta integral}) can be transformed
into a contour integral in the lower half-plane. The latter integral
contains contributions from all poles $\theta_{j}$ of $\alpha(\theta)$,
weighted by $e^{-2i\theta_{j}/\varepsilon}$. The dominant contribution
comes from the pole which is closest to the real line, say $\theta_{1}$.
Heuristically, this pole will be also the closest to the active epoch
around $\theta=\theta_{0}$. Hence\begin{equation}
\beta\sim\exp\left(-2\varepsilon^{-1}\textrm{Im}\left(\theta_{0}-\theta_{1}\right)\right),\end{equation}
which is of the same order as $\exp\left(-2n_{\max}(\theta_{0})\right)$.
The precision of the WKB approximation away from $\theta=\theta_{0}$
will be better than $\exp(-2n_{\max}(\theta_{0}))$, however, the
particle production will also decrease. We conclude that the uncertainty
inherent in the definition of particles during an {}``active epoch''
is quantitatively the same as the intensity of particle production
at that time.

\section{Examples \label{sec:Examples}}

In this section we shall investigate the precision of the WKB approximation
for particular functions $\omega(t)$. We use the formalism developed
in the previous sections and set $\varepsilon=1$.

The first example is\begin{equation}
\omega(t)=\omega_{0}\left(1+A\tanh\frac{t}{T}\right),\label{eq:omega ex 1}\end{equation}
where $\omega_{0},A$ and $T$ are constants. There are two super-adiabatic
regimes at $t\rightarrow\pm\infty$. Let us compute the optimal order
$n_{\max}$ for the WKB approximation $x_{\textrm{WKB}}^{(n)}(t_{0})$
at some intermediate time $t_{0}$. The function $\omega(t)$ has
zeros at \begin{equation}
t=t_{*}+i\pi kT,\quad k\in\mathbb{Z};\quad t_{*}\equiv\frac{i\pi}{2}T-\frac{T}{2}\ln\frac{1+A}{1-A}.\end{equation}
The closest zeros to a real point $t_{0}$ will be $\pm t_{*}$. Then
the optimal order of the WKB series is found as\begin{equation}
n_{\max}=\left|\int_{t_{0}}^{t_{*}}\omega(t)dt\right|=\omega_{0}\left|t_{*}-t_{0}+AT\ln\frac{\cosh\frac{t_{*}}{T}}{\cosh\frac{t_{0}}{T}}\right|,\label{eq:nmax 1}\end{equation}
which grows linearly with $t_{0}$ for $t_{0}\rightarrow\pm\infty$,
namely\begin{equation}
n_{\max}\approx\omega_{0}\left|t_{0}\right|\left(1\pm A\right)=\omega(t_{0})\left|t_{0}\right|,\quad t_{0}\rightarrow\pm\infty.\end{equation}
 The smallest term of the series is estimated as\begin{equation}
S_{n_{\max}}\sim\sqrt{\frac{\pi}{n_{\max}}}\exp\left(-2n_{\max}\right).\end{equation}
The worst precision is found near $t_{0}\approx0$ where we have (for
$A\ll1$)\begin{equation}
n_{\max}=\omega_{0}\left|t_{*}+AT\ln\cosh\frac{t_{*}}{T}\right|\approx\frac{\pi\omega_{0}T}{2}.\label{eq:n max worst 1}\end{equation}

\begin{figure}
\begin{center}\psfrag{t1}{1} \psfrag{t2}{2} \psfrag{t3}{3} \psfrag{t4}{4} \psfrag{t5}{5} \psfrag{t6}{6} \psfrag{t7}{7} \psfrag{t8}{8} \psfrag{sss}{$|S_n|/S_0$} \psfrag{nnn}{$n$}\includegraphics[%
  width=2.5in]{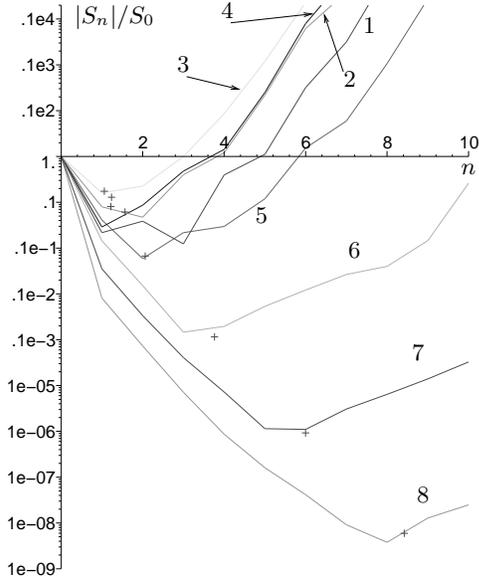}\end{center}

\caption{Magnitudes of first 10 terms $S_{n}$, $n=1,...,10$, of the WKB
series for $\omega(t)$ given by Eq.~(\ref{eq:omega ex 1}) and different
values of $t_{0}$. The values $S_{n}$ are normalized to $S_{0}$
and shown in logarithmic scale. The curves are labeled 1-8 according
to the consecutive values of $t_{0}$ in Table~\ref{cap:Numerical-values-table}.
Crosses indicate the error estimates; abscissas of crosses are set
to the predicted values of $n_{\max}(t_{0})$. The crosses are close
to the lowest-error terms on the corresponding curves, indicating
a good agreement between the predicted and the numerically obtained
values. \label{cap:Magnitudes-of-first}}
\end{figure}

For the purposes of numerical calculation, we chose $\omega_{0}=3$,
$A=2$, and $T=3/2$. The magnitudes of the first 10 terms $S_{1}$,
..., $S_{10}$ of the WKB series~(\ref{eq:W series}) are plotted
in Fig.~\ref{cap:Magnitudes-of-first}, together with the error estimates
(see Table~\ref{cap:Numerical-values-table} for numerical values).
It is clear from the plot that the terms $S_{n}$ indeed start to
grow after about $n=n_{\max}$, and that the error estimate agrees
with the numerically obtained smallest terms of the series.

\begin{table}
\begin{center}\begin{tabular}{|c|c|c|c|c|c|c|c|c|}
\hline 
$t_{0}$&
$-2.0$&
$-1.5$&
$-1.0$&
$-0.5$&
$0.0$&
$0.5$&
$1.0$&
$1.5$\tabularnewline
\hline 
$n_{\max}$&
$1.5$&
1.2&
1.0&
1.2&
2.0&
3.7&
6.0&
8.4\tabularnewline
\hline 
$\frac{S_{n_{\max}}}{S_{0}}$&
0.06&
0.12&
0.17&
0.08&
0.007&
$10^{-4}$&
$10^{-6}$&
6$\cdot$1$0^{-9}$\tabularnewline
\hline
\end{tabular}\end{center}

\caption{Numerical values of the predicted optimal truncation order $n_{\max}$
and error estimates $S_{n_{\max}}$ (see Fig.~\ref{cap:Magnitudes-of-first}).
The predicted values of $n_{\max}$ are not integer since they are
computed from Eq.~(\ref{eq:nmax 1}). \label{cap:Numerical-values-table}}
\end{table}

For comparison, let us approximately compute the particle occupation
number after the {}``in-out'' transition from $t=-\infty$ to $t=+\infty$.
The leading-order contribution to the particle number is proportional
to the term $u_{1}$ of the Bremmer series. It follows from Eq.~(\ref{eq:u2np1})
that \begin{equation}
\lim_{t\rightarrow+\infty}u_{1}(t)=\int_{-\infty}^{+\infty}dt\frac{\dot{\omega}}{2\omega}\exp\left(-2i\int_{0}^{t}\omega(t')dt'\right).\end{equation}
The integral can be evaluated as a sum of residues of the integrand
over its complex poles at \begin{equation}
t=t_{*}-i\pi kT,\quad k=0,1,2,...\end{equation}
Only the simple poles in the lower half-plane need to be accounted
for. (The function $\omega(t)$ also has poles at $t=\left(i\pi/2+i\pi k\right)T$,
which do not contribute to the integral.) After some algebra, the
result is found to be\begin{equation}
\lim_{t\rightarrow+\infty}u_{1}(t)=\frac{\pi}{2\sinh\left[\pi\omega_{0}T(1-A)\right]}.\end{equation}
Thus the particle number is of the same order as the worst precision
of the WKB series, $\sim\exp(-2n_{\max})$, where $n_{\max}$ is estimated
by Eq.~(\ref{eq:n max worst 1}) for $A\ll1$. This confirms the
general conclusions of Sec.~\ref{sub:Particle-production-and}.

The second example is\begin{equation}
\omega(t)=\omega_{0}\left(1+\frac{t^{4}}{T^{4}}\right).\label{eq:omega ex 2}\end{equation}
The four complex roots of $\omega(t)$ are\begin{equation}
t_{k}=T\exp\left(\frac{i\pi}{4}+\frac{i\pi k}{2}\right),\quad k=0,1,2,3.\end{equation}
The optimal truncation order is estimated as the smallest of the following
four numbers,\begin{equation}
n_{\max}=\min_{k}\left|\int_{t_{0}}^{t_{k}}\negmedspace\omega(t)dt\right|=\omega_{0}\min_{k}\left|t_{k}-t_{0}+\frac{\left(t_{k}-t_{0}\right)^{5}}{5T^{4}}\right|.\end{equation}
For $t_{0}>0$, the value of $n_{\max}$ is determined by the closest
roots $\pm T\exp(i\pi/4)$. For instance, at $t_{0}=0$ we have $n_{\max}\approx\frac{6}{5}\omega_{0}T$,
while for large $t_{0}\gg T$ we get $n_{\max}\approx\frac{1}{5}\omega_{0}t_{0}^{5}T^{-4}$.
The rapid growth of the allowed number of terms for $t_{0}\rightarrow\infty$
indicates a super-adiabatic regime.

\begin{figure}
\begin{center}\psfrag{t1}{1} \psfrag{t2}{2} \psfrag{t3}{3} \psfrag{t4}{4} \psfrag{t5}{5} \psfrag{t6}{6} \psfrag{t7}{7} \psfrag{t8}{8} \psfrag{sss}{$|S_n|/S_0$} \psfrag{nnn}{$n$}\includegraphics[%
  width=2.5in]{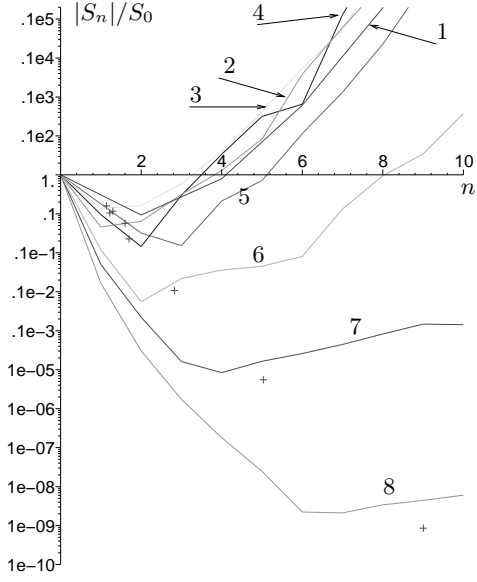}\end{center}

\caption{Analogous plot as in Fig.~\ref{cap:Magnitudes-of-first} for the
function $\omega(t)$ given by Eq.~(\ref{eq:omega ex 2}). The curves
are labeled 1-8 according to the consecutive values of $t_{0}$ in
Table~\ref{cap:Numerical-values-table2}. \label{cap:Magnitudes-of-2}}
\end{figure}

For the numerical calculation, we chose $\omega_{0}=1$ and $T=2$.
The results are shown in Fig.~\ref{cap:Magnitudes-of-2} and Table~\ref{cap:Numerical-values-table2}.

\begin{table}
\begin{center}\begin{tabular}{|c|c|c|c|c|c|c|c|c|}
\hline 
$t_{0}$&
0.0&
$0.5$&
$1.0$&
$1.5$&
$2.0$&
$2.5$&
$3.0$&
$3.5$\tabularnewline
\hline 
$n_{\max}$&
1.6&
1.3&
1.1&
1.2&
1.7&
2.8&
5.0&
9.0\tabularnewline
\hline 
$\frac{S_{n_{\max}}}{S_{0}}$&
0.06&
0.1&
0.16&
0.10&
0.02&
$10^{-3}$&
5$\cdot$1$0^{-6}$&
1$0^{-9}$\tabularnewline
\hline
\end{tabular}\end{center}

\caption{Numerical values of the predicted optimal truncation order and error
for Fig.~\ref{cap:Magnitudes-of-2}. \label{cap:Numerical-values-table2}}
\end{table}

\begin{figure}
\begin{center}\includegraphics[%
  width=2.5in]{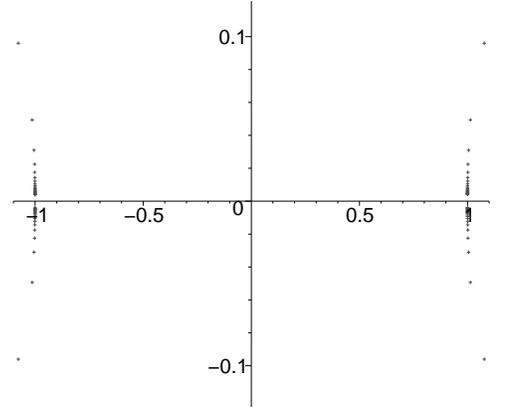}\end{center}

\caption{Roots of the function $\omega(t)$ given by Eq.~(\ref{eq:omega ex 3});
here $t$ is measured in units of $T$. The roots form sequences that
accumulate around $t=\pm T$ and curve away from $t=0$ if $\left|A\right|<1$.\label{cap:Roots-of-the}}
\end{figure}

The final example involves the function\begin{equation}
\omega(t)=\left\{ \begin{array}{l}
\omega_{0}\left[1+A\exp\left(-\frac{T^{2}}{T^{2}-t^{2}}\right)\right],\quad\left|t\right|<T;\\
\omega(t)=\omega_{0},\quad\left|t\right|\geq T,\end{array}\right.\label{eq:omega ex 3}\end{equation}
which exhibits super-adiabatic regimes at $t\leq-T$ and $t\geq T$.
We shall assume that $\left|A\right|\ll1$. The function $\omega(t)$
has zeros at\begin{equation}
t_{k}=\pm T\sqrt{1+\frac{1}{i\pi-\ln A+2i\pi k}},\quad k\in\mathbb{Z},\end{equation}
which form sequences converging to the essential singularities at
$t=\pm T$ (see Fig.~\ref{cap:Roots-of-the}). Thus the nearest singularity
to a point $t_{0}$ on the real line is effectively the point $t=T$.
Then the estimate~(\ref{eq:nmax def}) gives\begin{equation}
n_{\max}=\left|\int_{t_{0}}^{T}\omega(t)dt\right|\approx\omega_{0}\left(T-t_{0}\right),\quad0\leq t_{0}\leq T.\end{equation}
However, this estimate was derived only for functions with simple
isolated zeros, while in the present case there exist infinitely many
zeros in any neighborhood of $t=\pm T$. So we do not expect a close
agreement with the numerically obtained values of $n_{\max}$. 

\begin{figure}
\begin{center}\psfrag{t1}{1} \psfrag{t2}{2} \psfrag{t3}{3} \psfrag{t4}{4} \psfrag{t5}{5} \psfrag{t6}{6} \psfrag{t7}{7} \psfrag{sss}{$|S_n|/S_0$} \psfrag{nnn}{$n$}\includegraphics[%
  width=2.5in]{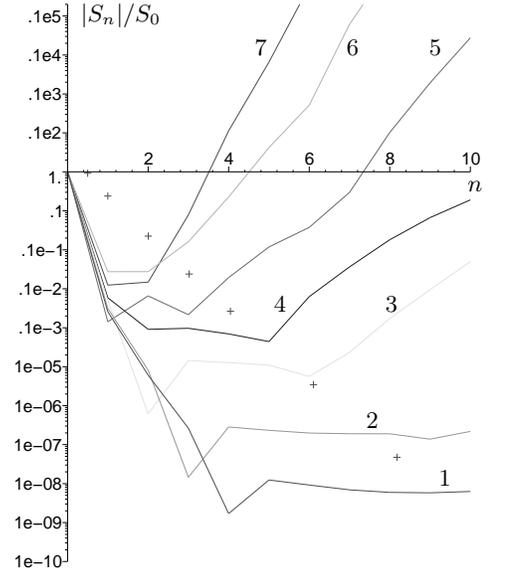}\end{center}

\caption{Analogous plot as in Fig.~\ref{cap:Magnitudes-of-first} for the
function $\omega(t)$ given by Eq.~(\ref{eq:omega ex 3}). The curves
are labeled 1-7 according to the consecutive values of $t_{0}$ in
Table~\ref{cap:Numerical-values-table3}. Relatively poor agreement
between crosses (the estimated values of $n_{\max}$) and the numerically
obtained values is due to the presence of essential singularities
in $\omega(t)$. \label{cap:Magnitudes-of-3}}
\end{figure}

For the numerical calculation, we chose $\omega_{0}=1$, $T=8$, and
$A=0.1$. The results are given in Fig.~\ref{cap:Magnitudes-of-3}
and Table~\ref{cap:Numerical-values-table3}. The computed values
show that a few first terms of the WKB series \emph{can} be used near
$t_{0}=T$; the estimate $n_{\max}\approx0$ is invalid in that regime.

\begin{table}
\begin{center}\begin{tabular}{|c|c|c|c|c|c|c|c|}
\hline 
$t_{0}$&
0.0&
$2.0$&
$4.0$&
$5.0$&
$6.0$&
$7.0$&
$7.5$\tabularnewline
\hline 
$n_{\max}$&
8.2&
6.1&
4.0&
3.0&
2.0&
1.0&
0.5\tabularnewline
\hline 
$\frac{S_{n_{\max}}}{S_{0}}$&
5$\cdot$1$0^{-8}$&
3$\cdot$1$0^{-6}$&
2$\cdot$1$0^{-4}$&
0.002&
0.02&
$0.24$&
0.9\tabularnewline
\hline
\end{tabular}\end{center}

\caption{Numerical values of the optimal truncation order and error for Fig.~\ref{cap:Magnitudes-of-3}.
\label{cap:Numerical-values-table3}}
\end{table}

\section{Summary}

In this paper I have reviewed some aspects of the adiabatic approximation
and its application to cosmological particle creation, and presented
new results. It is well known that there exists a fundamental limit
on the accuracy of the notion of particles in curved spacetimes, due
to Heisenberg uncertainty relations~\cite{BirDav82}. A quantitative
investigation of this accuracy is the main focus of the present paper.
I showed that the best attainable precision in the definition of particles
is exponentially small and of the same order as the typical particle
production expected during the same epoch. The conclusion is that
the ambiguity inherent in the definition of particles is precisely
due to the possibility of particle production. 

The main technical issue was to obtain an explicit estimate of the
highest attainable precision of the WKB approximation. I have demonstrated
that the WKB approximation involves a divergent series and derived
a novel formula~(\ref{eq:nmax def}) for the optimal truncation order
$n_{\max}$ of that series. I also estimated the error of the optimally
truncated WKB series {[}Eq.~(\ref{eq:error nmax}){]}. The error
is exponentially small since typical values of $n_{\max}$ will be
large if the adiabaticity condition~(\ref{eq:adiab cond}) holds.
Physically, the value of $n_{\max}$ is determined by the number of
oscillation periods during the time of appreciable change in the frequency
$\omega(t)$. Finally, I have presented analytic and numerical examples
illustrating the validity of these estimates.

\section*{Acknowledgments}

I am indebted to Gerald Dunne, Larry Ford, Stephen Fulling, Matthew
Parry, and Alex Vilenkin for helpful discussions.

\appendix

\section{Growth of derivatives of analytic functions\label{sec:Growth-of-derivatives}}

Here I derive some formulae used in Sec.~\ref{sub:Divergence-of-the}
and \ref{sec:Properties-of-the}.

By definition, an analytic function $w(\theta)$ can be represented
by a Taylor series at a regular point $\theta_{0}$,\begin{equation}
w(\theta_{0}+z)=\sum_{n=0}^{\infty}\frac{z^{n}}{n!}w^{(n)}(\theta_{0}),\label{eq:w Taylor series}\end{equation}
and it is well known that this series converges absolutely within
a circle $\left|z-\theta_{0}\right|<R_{1}$, where $R_{1}$ is the
distance between the point $\theta_{0}$ and the nearest singularity
of $w(\theta)$ in the complex $\theta$ plane. Suppose for simplicity
that the function $w(\theta)$ has simple poles at $\theta=\theta_{j}$
with residues $-c_{j}$, and that $\theta_{1}$ is the pole nearest
to $\theta_{0}$. Then we may express $w(\theta)$ as\begin{equation}
w(\theta)=\frac{c_{1}}{\theta_{1}-\theta}+w_{2}(\theta),\end{equation}
 where the auxiliary function $w_{2}(\theta)$ is analytic within
a larger circle $\left|z-\theta_{0}\right|<R_{2}$, $R_{2}>R_{1}$.
The $n$-th derivative of $w(\theta)$ is therefore\begin{equation}
w^{(n)}(\theta_{0})=\frac{n!c_{1}}{\left(\theta_{1}-\theta_{0}\right)^{n+1}}+w_{2}^{(n)}(\theta_{0}).\end{equation}
The same reasoning may be applied to the function $w_{2}(\theta)$
and the result is the following asymptotic estimate for the growth
of derivatives,\begin{align}
w^{(n)}(\theta_{0}) & =n!\left[\frac{c_{1}}{\left(\theta_{1}-\theta_{0}\right)^{n+1}}+\frac{c_{2}}{\left(\theta_{2}-\theta_{0}\right)^{n+1}}+...\right]\nonumber \\
 & \sim\frac{n!}{R_{1}^{n+1}}c_{1}\left[1+O\left(\left(R_{2}/R_{1}\right)^{n}\right)\right].\label{eq:w est 1}\end{align}

The estimate~(\ref{eq:w est 1}) is straightforwardly generalized
to the case when $w(\theta)$ has poles of higher order, e.g.\[
w(\theta)=\frac{b_{1}}{\left(\theta_{1}-\theta\right)^{2}}+\frac{c_{1}}{\theta_{1}-\theta}+w_{2}(\theta),\]
in which case we have\[
w^{(n)}(\theta_{0})\sim\frac{n!}{R_{1}^{n}}\left((n+1)\frac{b_{1}}{R_{1}}+c_{1}\right)\left[1+O\left(\left(R_{1}/R_{2}\right)^{n}\right)\right].\]
 Similar estimates can be easily obtained for the case when $w(\theta)$
has more than one pole at the same distance from the initial point
$\theta_{0}$.

\end{document}